\documentclass{aa}  
\usepackage{graphicx}
\usepackage{caption}
\usepackage{subcaption}
\usepackage{placeins}
\usepackage{booktabs}
\usepackage{stfloats}
\usepackage{xcolor}
\usepackage[normalem]{ulem}

\usepackage{orcidlink}

\usepackage{txfonts}
\usepackage{xspace}
\usepackage{gensymb}
\usepackage{hyperref}
\usepackage{lipsum}
\hypersetup{
    colorlinks=true,
    linkcolor=blue,
    filecolor=magenta,      
    urlcolor=blue,
    citecolor=blue,
}

\def\hess{H.E.S.S.\xspace}
\def\fermi{\textit{Fermi}-LAT\xspace}
\def\ssftt{SS~433\xspace}
\def\xray{X-ray\xspace}
\def\vsgr{V4641~Sgr\xspace}
\def\grs{GRS~1915+105\xspace}
\def\maxi{MAXI~J1820+070\xspace}

\defcitealias{Acharyya2026}{H26}

\begin{document} 

   \title{An \xray binary origin for the LHAASO peanut}

   \author{Laura Olivera-Nieto \inst{\ref{inst:API}, \ref{inst:GRAPPA}}\orcidlink{0000-0002-9105-0518}  
  \and Jakob~van~den~Eijnden\inst{\ref{inst:API}}\orcidlink{0000-0002-5686-0611}
    \and Rembrand Ruppert\inst{\ref{inst:API}\orcidlink{0009-0006-9448-2937}}
    \and Pikky Atri\inst{\ref{inst:ASTRON},\ref{inst:Radboud}}\orcidlink{0000-0001-8125-5619}
    \and Sera Markoff\inst{\ref{inst:API}, \ref{inst:GRAPPA}, \ref{inst:CAM}}\orcidlink{0000-0001-9564-0876}
   \and Kaya~Mori\inst{\ref{inst:Col}}\orcidlink{0000-0002-9709-5389}
   \and Gabriele~Ponti\inst{\ref{inst:INAF}, \ref{inst:CLAP},\ref{inst:MPE}}\orcidlink{0000-0003-0293-3608}
    \and Brian~Reville\inst{\ref{inst:MPIK}}\orcidlink{0000-0002-3778-1432} 
     \and Caterina~Tresoldi\inst{\ref{inst:INAF}, \ref{inst:CLAP}}\orcidlink{0009-0002-1992-0707}
    \and Naomi~Tsuji\inst{\ref{inst:ICRR}, \ref{inst:RIKEN}}\orcidlink{0000-0001-7209-9204}   }
    \institute{Anton Pannekoek Institute for Astronomy, University of Amsterdam, Science Park 904, 1098 XH Amsterdam, The Netherlands\label{inst:API}
    \and Gravitation and Astroparticle Physics Amsterdam Institute, University of Amsterdam, Science Park 904, 1098 XH Amsterdam, The Netherlands\label{inst:GRAPPA}
    \and ASTRON, Netherlands Institute for Radio Astronomy, Oude Hoogeveensedijk 4, 7991 PD Dwingeloo, The Netherlands\label{inst:ASTRON}
     \and Department of Astrophysics/IMAPP, Radboud University, P.O. Box 9010, 6500 GL Nijmegen, The Netherlands\label{inst:Radboud}
    \and Institute of Astronomy, University of Cambridge, Madingley Road, Cambridge CB3 0HA, United Kingdom  \label{inst:CAM}
        \and Columbia Astrophysics Laboratory, Columbia University, New York, NY 10027, USA \label{inst:Col}
        \and INAF–Osservatorio Astronomico di Brera, Via E.Bianchi 46, 23807, Merate (LC), Italy\label{inst:INAF}
    \and Como Lake Center for Astrophysics (CLAP), DiSAT, Università degli Studi dell’Insubria, viaValleggio 11, 22100 Como, Italy \label{inst:CLAP}
    \and Max-Planck-Institut f\"{u}r extraterrestrische Physik  (MPE), Giessenbachstrasse, Garching, 85748, Germany\label{inst:MPE}
    \and Max-Planck-Institut f\"{u}r Kernphysik (MPIK), Saupfercheckweg, Heidelberg D-69117, Germany\label{inst:MPIK}
       \and Institute for Cosmic Ray Research, The University of Tokyo, 5-1-5, Kashiwa-no-ha, Kashiwa, Chiba 277-8582, Japan\label{inst:ICRR}
    \and Interdisciplinary Theoretical \& Mathematical Science Center (iTHEMS), RIKEN, 2-1, Hirosawa, Wako, Saitama 351-0198, Japan\label{inst:RIKEN}
    }

   \date{Received XX XX, 2026; accepted XX XX, 2026}
\offprints{
\protect\\\email{\href{mailto:laura.olivera-nieto@mpi-hd.mpg.de}{l.oliveranieto@uva.nl}}
}
\abstract{The LHAASO "peanut" source is a remarkably extended ($\gtrsim$4$\degree$) ultra-high-energy gamma-ray emitter detected by the LHAASO observatory. While its angular extent and high-energy spectrum require a Galactic origin, its location at a Galactic latitude of $\approx$-17$\degree$ limits the possible known counterparts to a nearby millisecond pulsar. However, models accounting for the emission solely with this pulsar require the transport properties to operate near extreme parameter limits. We propose an alternative explanation where the peanut source is produced through the past activity of a now quiescent X-ray binary located at a distance of 1-2~kpc. We draw an analogy with known \xray binaries producing similar structures at high Galactic latitudes, and argue that the peanut emission could be explained by an almost identical physical scenario. We conclude that an \xray binary origin of the emission is plausible and predict that sufficiently sensitive \xray observations should be able to detect a quiescent \xray binary in the central region of the peanut source. If our hypothesis is confirmed, it would be the first time an \xray binary is identified in quiescence, opening a new avenue to detect such objects through their past gamma-ray emission.}


\maketitle


\section{Introduction}
\label{Intro}
The study of very-high-energy (VHE, TeV-scale) and ultra-high-energy (UHE, >100\,TeV-scale) gamma-ray sources has opened a new window onto the non-thermal universe. Wide-field instruments such as HAWC and LHAASO, and imaging atmospheric Cherenkov telescopes (IACTs) such as \hess, MAGIC and VERITAS have revealed a population of gamma-ray sources whose origins remain poorly understood. Despite extensive cataloguing attempts, the majority of VHE and UHE gamma-ray sources remain unassociated - that is, the astrophysical object responsible for the emission has not been identified~\citep[e.g.][]{HESSCollaboration2018, Cao2024a}. For the vast majority of unassociated sources this lack of confirmed counterpart is actually due to an over-abundance of potential candidates. Galactic gamma-ray sources are often extended, and since the distance to a gamma-ray source cannot be derived from its gamma-ray emission alone, it is often impossible to disentangle which astrophysical object is responsible for the emission in crowded Galactic environments. 

\begin{figure*}
	\centering
		\includegraphics[width=1\linewidth]{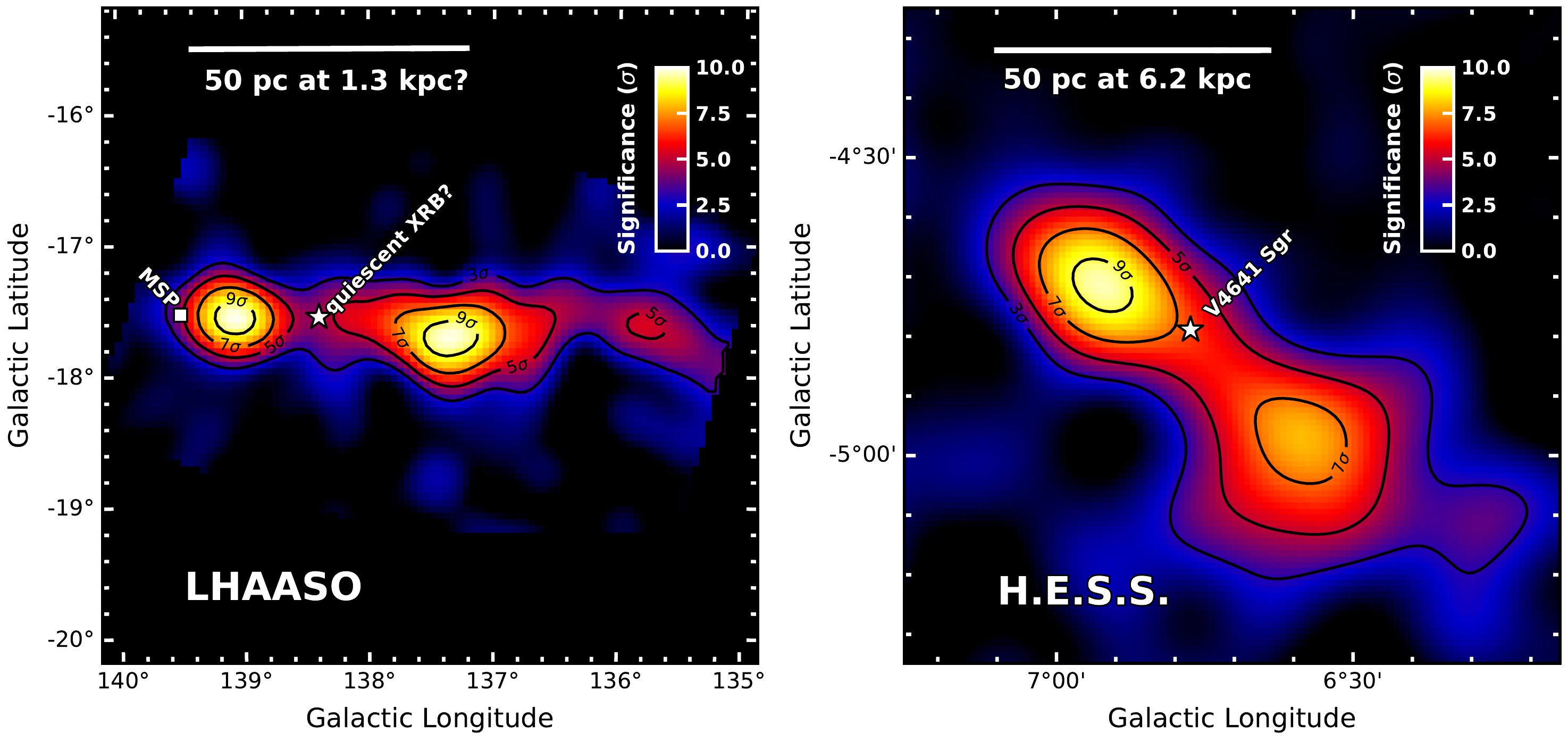} 
	\caption{\label{fig:maps} \textit{Left:} Significance map above 100~TeV of the peanut region from~\citet{Cao2025}. The position of the nearby MSP is indicated with the white square. An example position for the quiescent \xray binary hypothesized in this work is marked with a star. \textit{Right:} Significance map above 1~TeV of the \vsgr region from~\citet{Acharyya2026}. For visual consistency with the map provided by~\citet{Cao2025}, significance values below 0 are not shown here. The position of \vsgr is marked with a star.  }
\end{figure*}

There are, however, a handful of arguably more puzzling sources: those located at large Galactic latitudes. There, the reduced density of potential counterparts and the lower interstellar medium density mean that the source remains unassociated because no plausible counterpart is known. One such source is the so-called "peanut" source detected by LHAASO~\citep{Cao2025}. Spanning 0.45$\degree \times$4.6$\degree$ this large, asymmetric structure lies -17.5$\degree$ below the Galactic plane (see Figure~\ref{fig:maps}). The only 
known candidate counterparts are three pulsars. Two of them (PSR~J0211+4235 and PSR~J0209+4331) have spin-down energies on the order of $\dot{E}\sim 10^{30}$~erg/s, orders of magnitude lower than any pulsar detected in the gamma-ray band ($\dot{E}\gtrsim 10^{33}$~erg/s). The remaining object is a relatively powerful ($\dot{E}\sim 10^{35}$~erg/s) millisecond pulsar (MSP), MSP~J0218+4232.

This MSP is however located at the edge of the 4.5$\degree$ gamma-ray emission region (see Figure~\ref{fig:maps}), the latter being  misaligned with the MSP's proper motion. Consequently, if one adopts the distance to the MSP of around 3~kpc, a scenario where the MSP is the particle accelerator responsible for the peanut requires particles to travel in an extremely anisotropic manner for almost 250~pc. While this is in principle possible, it requires stringent assumptions on the nature of particle transport around the source~\citep[see e.g.][]{DeSarkar2026, Kleimenov2026}. Scenarios that invoke magnetic mirrors or anisotropic magnetic turbulence depend on specific, largely unconstrained properties of the ambient medium around MSP~J0218+4232. Additionally, VHE or UHE gamma-ray emission has never before been firmly associated with any other MSP~\citep[e.g.][]{HESSCollaboration2018, Cao2024a}.

In this letter, we propose an alternative scenario where the particle accelerator is located in the centre of the peanut region, a hypothesis also postulated by~\citet{DeSarkar2026}. Here, motivated by the similarity between the properties of the peanut and the gamma-ray emission discovered around some \xray binaries (XRBs), we explore a scenario where the central accelerator is a now quiescent \xray binary.

\section{\xray binaries}
\label{sec:xrbs}
\xray binaries are stellar binary systems hosting a single compact object. Accretion onto the compact object in these systems produces bright \xray radiation as matter flows inwards, which in turn can result in the launching of powerful, collimated outflows (`jets'), which are usually identified by their radio synchrotron emission. \xray emission from \xray binaries is variable and is characterized by periods of low activity ("quiescence") and outbursts, during which the source luminosity can rise by many orders of magnitude. There are a few hundred known Galactic \xray binaries: of those, around 20 are confirmed to host a black hole, with about eighty more candidates~\citep[e.g.][]{CorralSantana2016,Tetarenko2016,Olivera-Nieto2026}. 

\xray binaries spend most of their time in quiescence, up to 99\% in some estimates~\citep{Bailyn2017, Bahramian2020}. The sample of known \xray binaries is thus inevitably biased to those which have gone into outburst in the few decades in which \xray telescopes have been operating. Many \xray binaries have been observed in outburst only once, which implies the existence of objects that have not gone into outburst at all during this time~\citep{Tetarenko2016}. In fact, population synthesis estimates reveal that the number of \xray binaries currently in quiescence dwarfs the number of those that are active and thus detectable by current instruments~\citep[e.g.][]{Olejak2020}.


Due to their rapid variability in many other wavelengths, gamma-ray emission from XRBs was expected to be variable and correlate with outbursts or state transitions. However, recent results from HAWC and LHAASO~\citep[e.g.][]{Abeysekara2018, LhaasoCollaboration2025} have revealed that some XRBs power large ($\gtrsim$10-100~pc) gamma-ray nebulae, detected up to energies of several hundred TeV. The physical extent and luminosity of these gamma-ray structures imply they cannot form during a single outburst. Instead, they must reflect source activity integrated over timescales of at least thousands of years, far exceeding the several-decades baseline of direct transient observations. The natural implication of this discovery is that there might be "orphan" gamma-ray nebulae, powered in the past by an \xray binary which is now in quiescence~\citep[e.g.][]{Abaroa2026}. 

An example where such a situation would have taken place is \maxi; an \xray binary located at Galactic latitude $b=10\degree$. The \xray binary was first discovered in 2018, when it went into outburst~\citep{Kawamuro2018}. The LHAASO observatory has since reported the detection of a VHE gamma-ray source more than 10~pc away from the \xray binary, aligned with the jets~\citep{LhaasoCollaboration2025}. Because of the physical distance between the gamma-ray source and the binary, the former cannot possibly be related to the 2018 outburst and must have been present before. Consequently, had the LHAASO observatory come online a decade before it did in 2020, the gamma-ray source would have been discovered but have no known counterpart.

\subsection{Analogy with \vsgr}
\label{subsec:v4641}
\vsgr is a low-mass \xray binary hosting a compact object of mass $\mathrm{M}_{\mathrm{BH}}\approx$7~M$_{\odot}$~\citep{Goranskij2003, Orosz2001, MacDonald2014}. The companion star's mass is estimated to be $\mathrm{M}_{*}\approx$3~M$_{\odot}$, making it one of the most massive companions observed in an low-mass \xray binary system. \vsgr was first identified in 1999, when it underwent a strong outburst~\citep{Markwardt1999,Zand2000, Wijnands2000} during which its \xray luminosity increased by several orders of magnitude. A compact radio source was observed only briefly during this outburst~\citep{Hjellming2000}, revealing the presence of a relativistic jet~\citep{Marti2026}. Unlike other rapidly accreting, persistent \xray binaries (such as \ssftt or \grs), the only instance where the output of \vsgr was thought to have reached Eddington levels was during the 1999 outburst. Since then, the source has been comparatively quiet, with only smaller outbursts with average luminosities of $\lesssim 10^{36}$~erg~s$^{-1}$~\citep[e.g.][]{Tetarenko2016, Shaw2022}. Comparing the outburst history of \vsgr to other known \xray binaries reveals that, apart from a short period in 1999, its source output over the past few decades has remained below average for the population~\citep{Tetarenko2016}. 

\begin{figure}
	\centering
		\includegraphics[width=0.95\linewidth]{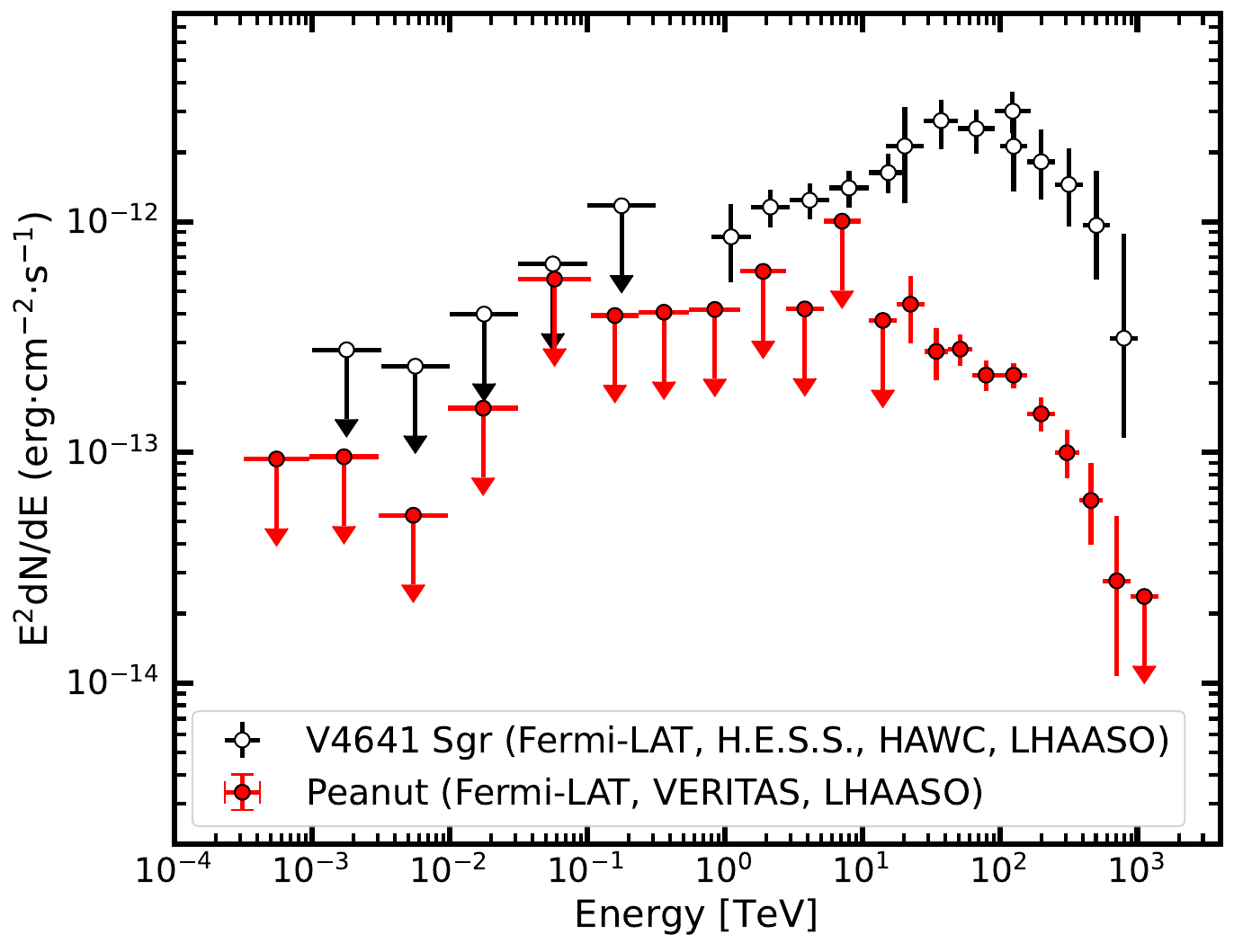} 
	\caption{\label{fig:flux} Spectral energy distribution of the peanut source (red points) as measured by LHAASO~\citep{Cao2025}, VERITAS and \fermi~\citep{VERITAS} compared to that of \vsgr (white points) as measured by \fermi~\citep{Zhao2025}, \hess~\citep{Acharyya2026} , HAWC~\citep{Alfaro2024}, and LHAASO~\citep{LhaasoCollaboration2025}. }
\end{figure}

In the gamma-ray band, this "average" \xray binary is revealed to be an extraordinary source. \vsgr produces a highly extended~\citep[$\sim1\degree$, ][]{Alfaro2024} asymmetrical nebula which can be resolved into two elliptical components~\citep{Acharyya2026} with indistinguishable spectral properties (see Figure~\ref{fig:maps}). The spectral energy distribution peaks at around 100~TeV and reaches photon energies of 800~TeV~\citep{LhaasoCollaboration2025}, requiring a remarkably efficient accelerator (see Figure~\ref{fig:flux}). The total luminosity above 1~TeV is $5\cdot10^{34}$~erg/s, an order of magnitude higher than the gamma-ray luminosity of \ssftt or \grs, both persistent and powerful \xray binaries\citep[e.g.][]{HESSCollaboration2024,LhaasoCollaboration2025}.

\vsgr is located more than 500~pc below the Galactic plane, in a region with no other objects that could account for the gamma-ray nebula and far beyond the scale of molecular gas in the Galaxy. Partly due to the lack of dense nearby gas, a leptonic origin is favoured as the emission mechanism responsible for most of the observed gamma-ray emission~\citep{Wan2026,Acharyya2026}. This interpretation is further supported by the recent report of \xray counterparts to the gamma-ray emission sites~\citep{Tsuji2026,Chi2026}. The combined \xray and gamma-ray observations require a relatively low magnetic field, of $B\lesssim$3~$\mu$G.

Detailed modelling of the emission around \vsgr has been carried out, among others, by~\citet{Wan2026} (first to propose a leptonic scenario),~\citet{Acharyya2026} (exploring the purely leptonic and purely hadronic cases) and~\citet{Kleimenov2026a} (considering  mixed lepto-hadronic scenarios). Regardless of the emission mechanism, the extent, flux and energy range of the gamma-ray emission require that the system must have been consistently more active in the past~\citep{Peretti2025, 2025ApJ...989L..25W}. While details depend on the scenario, the duration of this phase in the leptonic case with increased activity would need to be a few thousand years, while in the hadronic case,
it would require timescales on the order of millions of years~\citep[e.g.][]{Acharyya2026, Kleimenov2026a}.

At the distance of \vsgr~\citep[6.2~kpc, ][]{MacDonald2014}, the gamma-ray nebula has projected dimensions of around 100$\times$10~pc. The elongated structure is aligned with the direction of the jets in \vsgr~\citep{Marti2026}, suggesting a connection between them. In particular, the asymmetric morphology can be naturally explained by assuming that particle transport is dominated by advection in the jets with velocity $v_{\rm jet}$. Because of the relatively low magnetic field inferred from comparison with \xray observations, this flow only has to be mildly relativistic ($v_{\rm jet}$>0.1$c$)
for particles to cover dozens of pc without significant energy losses~\citep{Acharyya2026}.

\section{The peanut source}
\citet{Cao2025} report a highly extended morphology for the LHAASO peanut, with dimensions of 0.45$\degree \times$4.6$\degree$ (see Figure~\ref{fig:maps}). The emission can be separated into two round Gaussian components and a rectangular "strip" joining them, with indistinguishable spectra between the three. We note that the results of tests using a simpler hypothesis where the two components are allowed to be elliptical was not reported. The spectrum is well described by a power-law with an exponential cut-off, and extends up to energies of 700~TeV. A preliminary analysis of VERITAS observations~\citep{VERITAS} did not detect a lower-energy counterpart which places stringent limits on the 0.01-30~TeV range with additional \fermi data. No counterpart to the peanut is known in any other wavelength~\citep{Arias2026}, although we note that the coverage at such low Galactic latitudes can be much shallower than near the Galactic plane. Figure~\ref{fig:flux} shows the spectrum of the peanut compared to that of \vsgr. Similarly to the case of \vsgr, the lack of dense gas nearby favours a scenario where the majority of the emission is produced by electrons.

While~\citet{Cao2025} assume a physical distance to the peanut of 3~kpc to match that of the MSP, we note that the gamma-ray data alone provide no information on the distance beyond identifying the peanut as a not-too-distant Galactic source. At that distance, the structure would span almost 250~pc, forcing models to invoke fast and extremely anisotropic transport to explain it~\citep{Cao2025,DeSarkar2026,Kleimenov2026}.

In this work, we draw a parallel with \vsgr and instead consider the possibility that the peanut was produced in one or more phases of activity from a now quiescent \xray binary located somewhere in its central region. The lack of evidence for a massive star in the region of the potential \xray binary, as well as the spatial distributions of known low and high-mass \xray binaries, strongly disfavours a high-mass X-ray binary (see Appendix~\ref{app:opt}). A low-mass \xray binary scenario is also consistent with the necessary transient nature of source. 

Assuming the physical size of the peanut to be similar to nebulae detected near other \xray binaries would place it at a distance of 1 to 2~kpc. In the following, we will adopt a distance of 1.3~kpc, which corresponds to a 100~pc physical extent for the peanut but note that the real distance could be closer or farther, and the numbers that follow could be adjusted accordingly. The left panel of Figure~\ref{fig:spec} compares the luminosity of the peanut at this assumed distance. As can be seen in the figure, the spectral shape is consistent, with the peanut being roughly 300 times less luminous than \vsgr. Such a difference in luminosity is not unexpected: the other gamma-ray nebulae detected around \xray binaries have luminosities as low as 4 orders of magnitude less than \vsgr, which stands out as the brightest example. 

\section{A naive model of advection within a jet}
\label{sec:model}

\begin{figure*}
    \centering
    \begin{subfigure}{0.47\textwidth}
        \centering
        \includegraphics[width=0.97\textwidth]{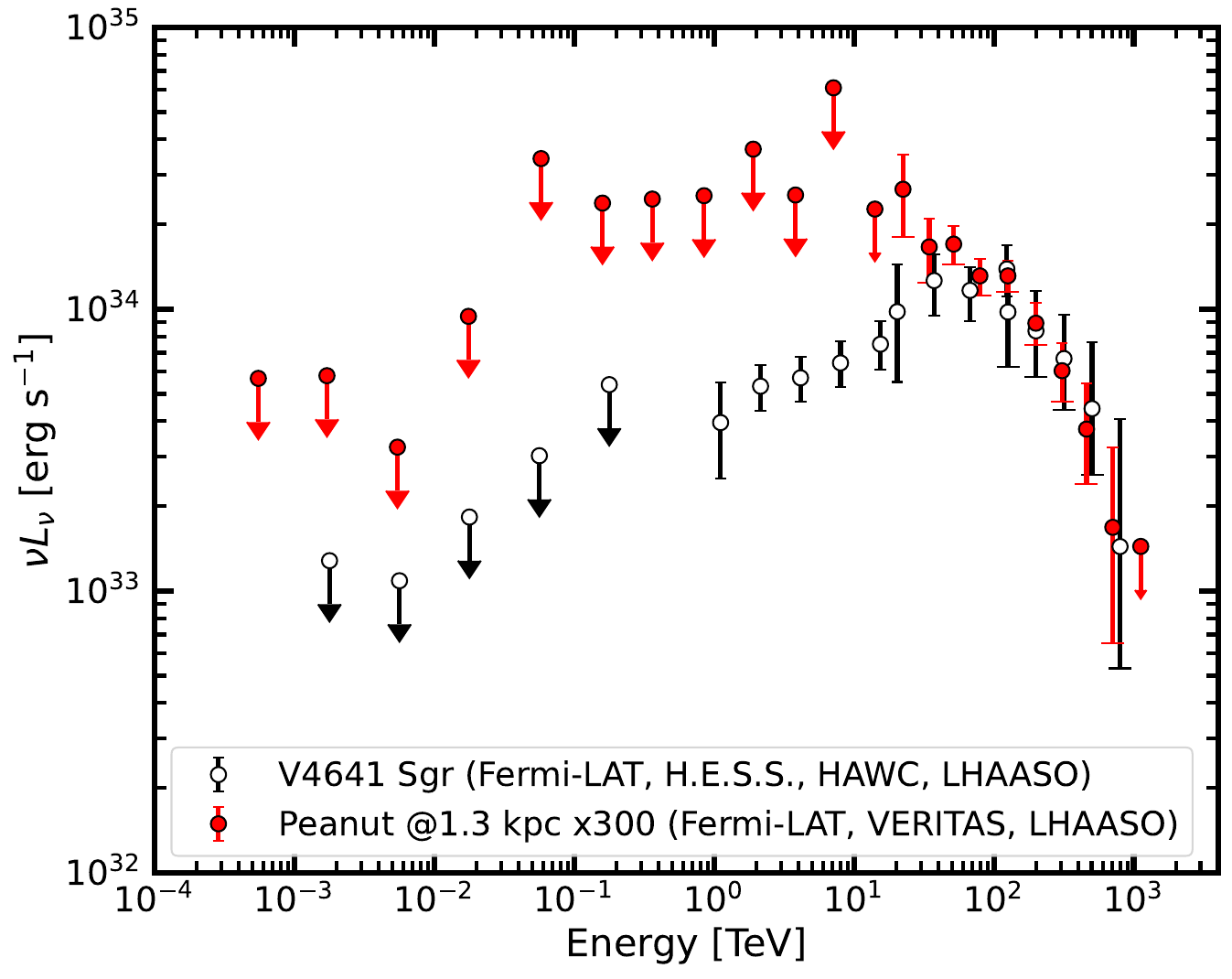} 
    \end{subfigure}
    \begin{subfigure}{0.47\textwidth}
        \centering
		\includegraphics[width=0.97\textwidth]{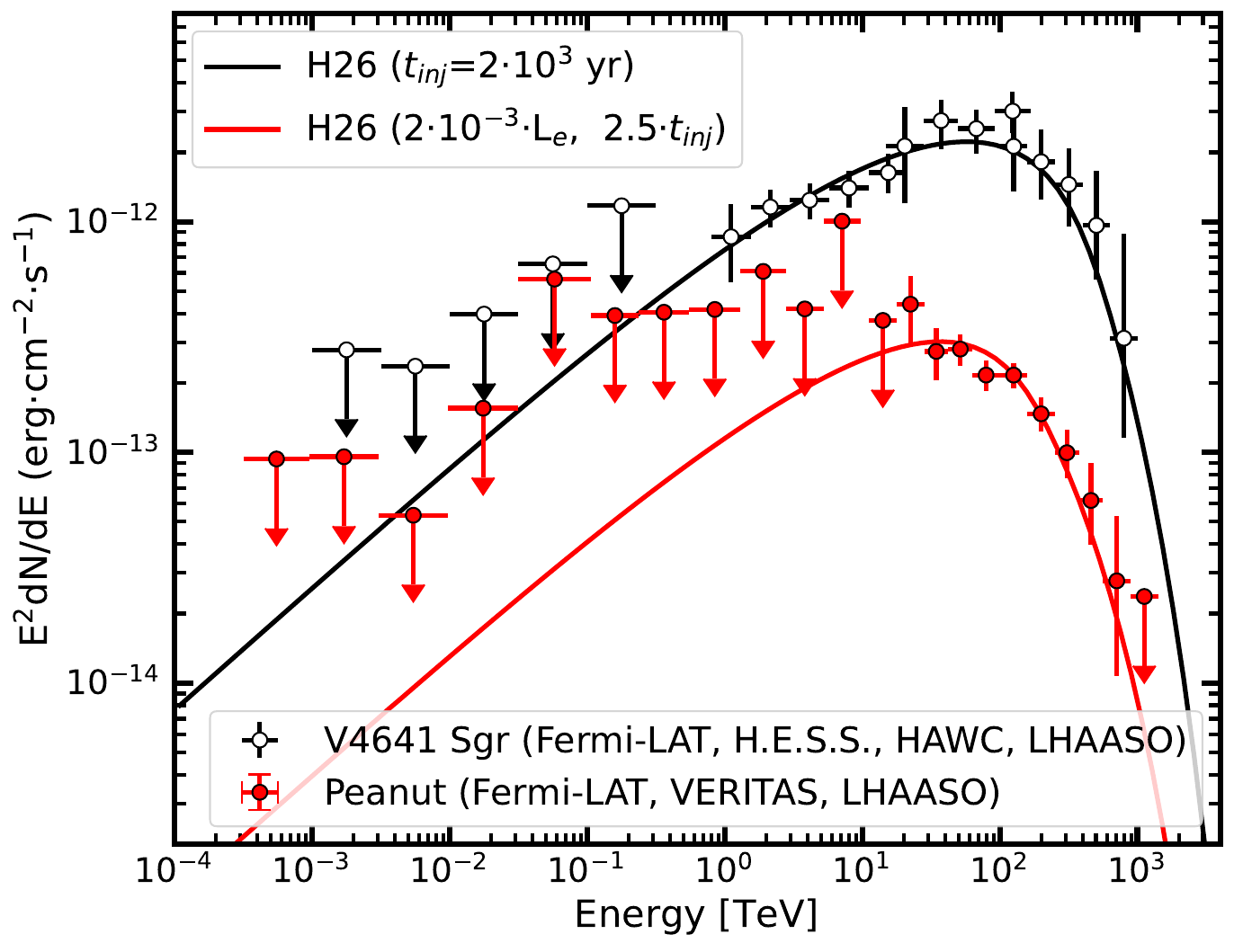} 
    \end{subfigure}\\

    \caption{\label{fig:spec}\textit{Left:} Symbols follow those of Figure~\ref{fig:flux} but with the flux converted to luminosity, adopting a distance of 6.2~kpc~\citep{MacDonald2014} for \vsgr and 1.3~kpc for the peanut, motivated by their relative sizes. The peanut luminosity has additionally been scaled by a factor of 300 to ease comparison. \textit{Right:} Flux measurements shown in Figure~\ref{fig:flux} with one of the models presented by~\citet{Acharyya2026} (black line) slightly adapted to describe the peanut spectrum (red line).}
\end{figure*}

Assuming a distance of 1.3~kpc, the peanut would be located approximately 370~pc below the Galactic plane, a comparable environment to that of \vsgr. For this reason, we will rely on a model developed for \vsgr in the following discussion. In particular, we adopt the leptonic, single-zone model presented in~\citet{Acharyya2026} (hereafter~\citetalias{Acharyya2026}). The model uses the  open-source GAMERA package~\citep{Hahn2015, Hahn2022}, which includes Klein-Nishina corrections for the IC scattering calculations~\cite[e.g.][]{Blumenthal1970}, to model both the temporal evolution of the particle distributions and the resulting radiation. The adopted ambient radiation field is a combination of a publicly available axisymmetric Galactic model for the energy densities of the diffuse interstellar radiation field~\citep{Popescu2017} and the Cosmic Microwave Background. 

Rather unsurprisingly, given the simplicity of this model, we find that minimal tweaks can reproduce the observed properties of the peanut. The right panel of Figure~\ref{fig:spec} shows an example where the~\citetalias{Acharyya2026} model for an injection time ($t_{\rm inj}$) of $2\cdot 10^3$~yr provides a good fit to the data by simply scaling the luminosity of injected electrons (L$_e$) down by a factor $\approx$300 (motivated by the gamma-ray luminosity comparison) and increasing $t_{\rm inj}$ by a factor of 2.5 to capture the turnover in the LHAASO spectrum towards lower energies. The full parameter set is collected in Table~\ref{tab:params}. We stress that this particular parameter set is simply an example, chosen to highlight the fact that the spectral properties of the peanut are similar to those of \xray binary nebulae. As was the case for the~\citetalias{Acharyya2026} model, different combinations of spectral index and $t_{\rm inj}$ would also provide a good fit to the data. 

The spatial properties of the peanut can be explained with a model analogous to that of \vsgr, as the similarity highlighted in Figure~\ref{fig:maps} suggests. We propose that the morphology of the peanut can be explained by two-sided advection by a (semi-) relativistic jet. This explanation naturally produces asymmetric morphology and the fast, energy-independent transport significantly simplifies the requirements on the model. Any leptonic scenario which does not invoke several re-acceleration sites requires the average magnetic field of the peanut to be low, in order to allow electrons to travel with minimal losses. \citet{Cao2025,Kleimenov2026} and~\citet{DeSarkar2026} all adopt B$\approx$1~$\mu$G. For such a magnetic field, the cooling time of a 1~PeV electron is around 10$^4$~yr. Consequently, even a relatively slow jet ($v_{\rm jet}\sim$0.1$c$) would be able to transport particles for hundreds of pc with minimal losses.

Note that the relaxed constraint on $v_{\rm jet}$ implies that our proposal does not require every property of the peanut nebula to match that of \vsgr. A similar physical model would naturally explain the data even if, e.g., the magnetic field were higher or the spatial extent different, by simply adjusting $v_{\rm jet}$. Jets from \xray binaries have been observed to display a wide range of velocities~\citep[see e.g. ][for a recent summary]{Fender2025}, and in a low-density environment like that of the peanut, they could propagate even farther.

\subsection{Model predictions}
\label{subsec:xray}
Our model makes one concrete prediction that should be verifiable with current and future observational facilities: the existence of a quiescent low-mass \xray binary somewhere between the two LHAASO hotspots. Other predictions, such as the existence of an extended, low-surface-brightness \xray counterpart, are not specific to the \xray binary interpretation and simply arise from assuming a leptonic nature for the emission, so we will not discuss them in detail here.

 Everything known about \xray binaries in quiescence has been learned by observing known systems when they are not in outburst. A handful of black-hole \xray binaries have been detected in quiescence. A0620-00 is the closest known \xray binary, located at a distance of $\sim1$~kpc. Because of that, its behaviour in deep quiescence has been characterized in detail, down to luminosities as low as 10$^{-8}$ times its Eddington luminosity~\citep{Gallo2006, Dincer2018} without the need for deep exposures. Another relatively nearby system, XTE J1118+480, is located at a distance of $\sim1.7$~kpc and has been observed down to even lower Eddington ratios thanks to its advantageous location at $b\approx62\degree$,  which significantly reduces the impact of absorption~\citep{McClintock2003}. Another high-latitude ($b\approx50\degree$) system, Swift~J1357.2-0933, has been detected to similarly low Eddington ratios despite its higher distance (2.3-6.3~kpc) because of the low absorption in its direction~\citep{Plotkin2016}. Other \xray binaries with higher (although often less accurate) distance estimates and less advantageous Galactic coordinates have also been detected in quiescence, with Eddington ratios ranging between 10$^{-8}$ and 10$^{-5}$~\citep[e.g.][]{Hameury2003,Hynes2009}. In particular, the farthest black hole \xray binary detected in quiescence is BW~Cir, located at $>$7~kpc. It has been detected down to Eddington ratios of $\lesssim$10$^{-6}$~\citep{Plotkin2021}.
 
 Because of the location of the peanut at $b\approx -17\degree$ and the assumed nearby location ($\lesssim$2~kpc), the quiescent \xray binary should be detectable with, e.g., deep Chandra or XMM observations. Scaling the lowest flux measured from A0620-00 in quiescence~\citep{Dincer2018} to a slightly larger distance results in a flux in the 2-10~keV range of around 6$\cdot 10^{-15}$~erg/cm$^2$/s, which could be reached with less than 100~ks of observation with the current generation of instruments. Planned and proposed \xray observatories such as NewAthena~\citep{Cruise2025} or the HEROIX telescope should be able to detect it without the need for such deep observations. Assuming the well-established radio-\xray correlation for black-hole systems~\citep[e.g.][]{Merloni2003, Falcke2004, Plotkin2012, Crook-Mansour2026}, the corresponding radio flux at 1.28~GHz would be $\sim$20$-$40~$\mu$Jy~\citep[similar to A0620-00 scaled to $1.3$ kpc, ][]{Gallo2006}, again faint but detectable with current instruments. If the compact object was instead a quiescent neutron star, it would not be detected as a radio source~\citep{Gallo2018}.


\section{Conclusions}
\label{sec:the_conclusions}
 We conclude that an \xray binary origin of the emission is plausible and predict that sufficiently sensitive \xray and radio observations should be able to detect a quiescent \xray binary in the central region of the peanut source. We predict that the jet position angle of this binary would align with the peanut major axis. If our hypothesis is confirmed, it would be the first time an \xray binary is identified in quiescence, opening a new avenue to detect such objects through their past gamma-ray emission. Additionally, a detection of gamma-ray emission from a dormant \xray binary would lend further support to the growing evidence supporting jetted \xray binaries as the source of cosmic rays above PeV energies~\citep[e.g.][]{2025ApJ...989L..25W, Kaci2025, Abaroa2026}.

\begin{acknowledgements}
LON acknowledges support from the Max-Planck Society and, with S.M., the European Research Council Synergy Grant ‘Blackholistic’ (grant agreement no. 101071643). PA is supported by the WISE fellowship program, which is financed by NWO. G.P. acknowledges financial support from the European Research Council (ERC) under the European Union’s Horizon 2020 research and innovation program HotMilk (grant agreement No. 865637).
\end{acknowledgements}

%
%

\bibliographystyle{aa}
\bibliography{uQ}

@ARTICLE{2001A&A...378..907R,
       author = {{Ramspeck}, M. and {Heber}, U. and {Moehler}, S.},
        title = "{Early type stars at high galactic latitudes. I. Ten young massive B-type stars}",
      journal = {\aap},
         year = 2001,
        month = nov,
       volume = {378},
        pages = {907-917},
          doi = {10.1051/0004-6361:20011246},
archivePrefix = {arXiv},
       eprint = {astro-ph/0109472},
 primaryClass = {astro-ph},
       adsurl = {https://scixplorer.org/abs/2001A&A...378..907R}
}

@ARTICLE{2023A&A...675A.199A,
       author = {{Avakyan}, A. and {Neumann}, M. and {Zainab}, A. and {Doroshenko}, V. and {Wilms}, J. and {Santangelo}, A.},
        title = "{XRBcats: Galactic low-mass X-ray binary catalogue}",
      journal = {\aap},
         year = 2023,
        month = jul,
       volume = {675},
          eid = {A199},
        pages = {A199},
          doi = {10.1051/0004-6361/202346522},
archivePrefix = {arXiv},
       eprint = {2303.16168},
 primaryClass = {astro-ph.HE},
       adsurl = {https://scixplorer.org/abs/2023A&A...675A.199A}
}

@ARTICLE{2023A&A...677A.134N,
       author = {{Neumann}, M. and {Avakyan}, A. and {Doroshenko}, V. and {Santangelo}, A.},
        title = "{XRBcats: Galactic High Mass X-ray Binary Catalogue★}",
      journal = {\aap},
         year = 2023,
        month = sep,
       volume = {677},
          eid = {A134},
        pages = {A134},
          doi = {10.1051/0004-6361/202245728},
archivePrefix = {arXiv},
       eprint = {2303.16137},
 primaryClass = {astro-ph.HE},
       adsurl = {https://scixplorer.org/abs/2023A&A...677A.134N}
}

@ARTICLE{2024A&A...681A..21S,
       author = {{Stoop}, M. and {Derkink}, A. and {Kaper}, L. and {de Koter}, A. and {Rogers}, C. and {Ram{\'\i}rez-Tannus}, M.~C. and {Guo}, D. and {Azatyan}, N.},
        title = "{The early evolution of young massive clusters. II. The kinematic history of NGC 6618/M 17}",
      journal = {\aap},
         year = 2024,
        month = jan,
       volume = {681},
          eid = {A21},
        pages = {A21},
          doi = {10.1051/0004-6361/202347383},
archivePrefix = {arXiv},
       eprint = {2311.04174},
 primaryClass = {astro-ph.GA},
       adsurl = {https://scixplorer.org/abs/2024A&A...681A..21S}
}

@ARTICLE{2025MNRAS.543...63P,
       author = {{Pantaleoni Gonz{\'a}lez}, M. and {Ma{\'\i}z Apell{\'a}niz}, J. and {Barb{\'a}}, R.~H. and {Reed}, B. Cameron and {Berlanas}, S.~R. and {Parras Rico}, A. and {Bodaghee}, A.},
        title = "{The Alma catalogue of OB stars ─ III. A cross-match with Gaia DR3 and an extension based on new spectral classifications}",
      journal = {\mnras},
         year = 2025,
        month = oct,
       volume = {543},
       number = {1},
        pages = {63-82},
          doi = {10.1093/mnras/staf1409},
archivePrefix = {arXiv},
       eprint = {2508.14875},
 primaryClass = {astro-ph.SR},
       adsurl = {https://scixplorer.org/abs/2025MNRAS.543...63P}
}

@ARTICLE{2019ApJ...887...93G,
       author = {{Green}, Gregory M. and {Schlafly}, Edward and {Zucker}, Catherine and {Speagle}, Joshua S. and {Finkbeiner}, Douglas},
        title = "{A 3D Dust Map Based on Gaia, Pan-STARRS 1, and 2MASS}",
      journal = {\apj},
         year = 2019,
        month = dec,
       volume = {887},
       number = {1},
          eid = {93},
        pages = {93},
          doi = {10.3847/1538-4357/ab5362},
archivePrefix = {arXiv},
       eprint = {1905.02734},
 primaryClass = {astro-ph.GA},
       adsurl = {https://scixplorer.org/abs/2019ApJ...887...93G}
}

@ARTICLE{2018JOSS....3..695M,
       author = {{Green}, {Gregory M.}},
        title = "{dustmaps: A Python interface for maps of interstellar dust}",
      journal = {The Journal of Open Source Software},
         year = "2018",
        month = "Jun",
       volume = {3},
       number = {26},
        pages = {695},
          doi = {10.21105/joss.00695},
       adsurl = {https://ui.adsabs.harvard.edu/abs/2018JOSS....3..695G}
}

@Article{Tetarenko2016,
  author        = {Tetarenko, B. E. and Sivakoff, G. R. and Heinke, C. O. and Gladstone, J. C.},
  journal       = {\apjs},
  title         = {WATCHDOG: A Comprehensive All-sky Database of Galactic Black Hole X-ray Binaries},
  year          = {2016},
  month         = feb,
  number        = {2},
  pages         = {15},
  volume        = {222},
  archiveprefix = {arXiv},
  doi           = {10.3847/0067-0049/222/2/15},
  eid           = {15},
  eprint        = {1512.00778},
  primaryclass  = {astro-ph.HE},
  url           = {https://ui.adsabs.harvard.edu/abs/2016ApJS..222...15T},
}

@Article{CorralSantana2016,
  author        = {Corral-Santana, J. M. and Casares, J. and Mu{\~n}oz-Darias, T. and Bauer, F. E. and Mart{\'\i}nez-Pais, I. G. and Russell, D. M.},
  journal       = {\aap},
  title         = {BlackCAT: A catalogue of stellar-mass black holes in X-ray transients},
  year          = {2016},
  month         = mar,
  pages         = {A61},
  volume        = {587},
  archiveprefix = {arXiv},
  doi           = {10.1051/0004-6361/201527130},
  eid           = {A61},
  eprint        = {1510.08869},
  primaryclass  = {astro-ph.HE},
  url           = {https://ui.adsabs.harvard.edu/abs/2016A&A...587A..61C},
}

@Article{Gallo2006,
  author        = {Gallo, E. and Fender, R. P. and Miller-Jones, J. C. A. and Merloni, A. and Jonker, P. G. and Heinz, S. and Maccarone, T. J. and van der Klis, M.},
  journal       = {\mnras},
  title         = {A radio-emitting outflow in the quiescent state of A0620-00: implications for modelling low-luminosity black hole binaries},
  year          = {2006},
  month         = aug,
  number        = {3},
  pages         = {1351-1360},
  volume        = {370},
  archiveprefix = {arXiv},
  doi           = {10.1111/j.1365-2966.2006.10560.x},
  eprint        = {astro-ph/0605376},
  primaryclass  = {astro-ph},
  url           = {https://ui.adsabs.harvard.edu/abs/2006MNRAS.370.1351G},
}

@Article{Plotkin2021,
  author        = {Plotkin, R. M. and Bahramian, A. and Miller-Jones, J. C. A. and Reynolds, M. T. and Atri, P. and Maccarone, T. J. and Shaw, A. W. and Gandhi, P.},
  journal       = {\mnras},
  title         = {Towards a larger sample of radio jets from quiescent black hole X-ray binaries},
  year          = {2021},
  month         = may,
  number        = {3},
  pages         = {3784-3795},
  volume        = {503},
  archiveprefix = {arXiv},
  doi           = {10.1093/mnras/stab644},
  eprint        = {2103.02178},
  primaryclass  = {astro-ph.HE},
  url           = {https://ui.adsabs.harvard.edu/abs/2021MNRAS.503.3784P},
}

@Article{Plotkin2016,
  author        = {Plotkin, Richard M. and Gallo, Elena and Jonker, Peter G. and Miller-Jones, James C. A. and Homan, Jeroen and Mu{\~n}oz-Darias, Teo and Markoff, Sera and Armas Padilla, Montserrat and Fender, Rob and Rushton, Anthony P. and Russell, David M. and Torres, Manuel A. P.},
  journal       = {\mnras},
  title         = {A clean sightline to quiescence: multiwavelength observations of the high Galactic latitude black hole X-ray binary Swift J1357.2-0933},
  year          = {2016},
  month         = mar,
  number        = {3},
  pages         = {2707-2716},
  volume        = {456},
  archiveprefix = {arXiv},
  doi           = {10.1093/mnras/stv2861},
  eprint        = {1512.01941},
  primaryclass  = {astro-ph.HE},
  url           = {https://ui.adsabs.harvard.edu/abs/2016MNRAS.456.2707P},
}

@Article{Marti2026,
  author        = {Mart{\'\i}, Josep and Luque-Escamilla, Pedro L.},
  journal       = {\mnras},
  title         = {Alignment of radio jets in the microquasar V4641 Sagittarii with its high-energy structures},
  year          = {2026},
  month         = jan,
  number        = {3},
  pages         = {staf2104},
  volume        = {545},
  archiveprefix = {arXiv},
  doi           = {10.1093/mnras/staf2104},
  eid           = {staf2104},
  eprint        = {2511.19695},
  primaryclass  = {astro-ph.HE},
  url           = {https://ui.adsabs.harvard.edu/abs/2026MNRAS.545f2104M},
}

@Article{LhaasoCollaboration2025,
  author        = {{LHAASO Collaboration} and Cao, Zhen and Aharonian, Felix and Bai, Yun-Xiang and Bao, Yi-Wei and Bastieri, Denis and Bi, Xiao-Jun and Bi, Yu-Jiang and Bian, Wen-Yi and Bukevich, Anatoly V. and Cai, Chengmiao and Cao, Wen-Yu and Cao, Zhe and Chang, Jin and Chang, Jin-Fan and Chen, Aming and Chen, En-Sheng and Chen, Guohai and Chen, Hua-Xi and Chen, Liang and Chen, Long and Chen, Ming-Jun and Chen, Ma-Li and Chen, Qi-Hui and Chen, Shi and Chen, Su-Hong and Chen, Song-Zhan and Chen, Tian-Lu and Chen, Xiao-Bin and Chen, Xuejian and Chen, Yang and Cheng, Ning and Cheng, Yao-Dong and Chung Chu, Ming and Cui, Ming-Yang and Cui, Shu-Wang and Cui, Xiao-Hong and Cui, Yi-Dong and Dai, Ben-Zhong and Dai, Hong-Liang and Dai, Zigao and Luobu, Danzeng and Diao, Yang-Xuan and Dong, Xu-Qiang and Duan, Kai-Kai and Fan, Jun-Hui and Fan, Yi-Zhong and Fang, Jun and Fang, Jian-Hua and Fang, Kun and Feng, Cun-Feng and Feng, Hua and Feng, Li and Feng, Shaohui and Feng, Xiao-Ting and Feng, Yi and Feng, You-Liang and Gabici, Stefano and Gao, Bo and Gao, Chuan-Dong and Gao, Qi and Gao, Wei and Gao, Wei-Kang and Ge, Maomao and Ge, Ting-Ting and Geng, Lisi and Giacinti, Gwenael and Gong, Guanghua and Gou, Quanbu and Gu, Min-Hao and Guo, Fu-Lai and Guo, Jing and Guo, Xiao-Lei and Guo, Yi-Qing and Guo, Ying-Ying and Han, Yi-Ang and Hannuksela, Otto A. and Hasan, Mariam and He, Hui-Hai and He, Hao-Ning and He, Jia-Yin and He, Xinyu and He, Yu and Hern{\'a}ndez-Cadena, Sergio and Hou, Bo-Wen and Hou, Chao and Hou, Xian and Hu, Hong-Bo and Hu, Shi-Cong and Huang, Chen and Huang, Dai-Hui and Huang, Jiajun and Huang, Tian-Qi and Huang, Wen-Jun and Huang, Xing-Tao and Huang, Xiao-Yuan and Huang, Yong and Huang, Yi-Yun and Ji, Xiao-Lu and Jia, Huan-Yu and Jia, Kang and Jiang, Hou-Bing and Jiang, Kun and Jiang, Xiao-Wei and Jiang, Ze-Jun and Jin, Min and Kaci, Samy and Kang, Ming-Ming and Karpikov, Ivan and Khangulyan, Dmitry and Kuleshov, Denis and Kurinov, Kirill and Li, Bing-Bing and Li, Cheng and Li, Cong and Li, Dan and Li, Fei and Li, Haibo and Li, Huicai and Li, Jian and Li, Jie and Li, Kai and Li, Long and Li, Rong-Lan and Li, Si-Da and Li, Tian-Yang and Li, Wen-Lian and Li, Xiu-Rong and Li, Xin and Li, Yuan and Li, Yizhuo and Li, Zhe and Li, Zhuo and Liang, En-Wei and Liang, Yun-Feng and Lin, Su-Jie and Liu, Bing and Liu, Cheng and Liu, Dong and Liu, Dang-Bo and Liu, Hu and Liu, Hai-Dong and Liu, Jia and Liu, Jia-Li and Liu, Ji-Ren and Liu, Mao-Yuan and Liu, Ruo-Yu and Liu, Si-Ming and Liu, Wei and Liu, X. and Liu, Yi and Liu, Yu and Liu, Yi-Nong and Lou, Yu-Qing and Luo, Qing and Luo, Yu and Lv, Hong-Kui and Ma, Bo-Qiang and Ma, Ling-Ling and Ma, Xin-Hua and Mao, Ji-Rong and Min, Zhen and Mitthumsiri, Warit and Mou, Guo-Bin and Mu, Hui-Jun and Neronov, Andrii and Ng, Kenny Chun Yu and Ni, Ming-Yang and Nie, Lin and Ou, Le-Jian and Pattarakijwanich, Petchara and Pei, Zhi-Yuan and Qi, Jin-Can and Qi, Meng-Yao and Qin, Jia-Jun and Raza, Ali and Ren, Chong-Yang and Ruffolo, David and S{\'a}iz, Alejandro and Semikoz, Dmitri and Shao, Lang and Shchegolev, Oleg and Shen, Yun-Zhi and Sheng, Xiang-Dong and Shi, Zhaodong and Shu, Fu-Wen and Song, Hui-Chao and Stenkin, Yuri V. and Stepanov, Vladimir and Su, Yang and Sun, Dongxu and Sun, Hao and Sun, Qinning and Sun, Xiaona and Sun, Zhibin and Hussain Tabasam, Nabeel and Takata, Jumpei and Tam, Pak Hin Thomas and Tan, Hong-Bin and Tang, Qingwen},
  journal       = {National Science Review},
  title         = {Ultrahigh-Energy Gamma-ray Emission Associated with Black Hole-Jet Systems},
  year          = {2025},
  month         = dec,
  number        = {12},
  pages         = {nwaf496},
  volume        = {12},
  archiveprefix = {arXiv},
  doi           = {10.1093/nsr/nwaf496},
  eid           = {nwaf496},
  eprint        = {2410.08988},
  primaryclass  = {astro-ph.HE},
  url           = {https://ui.adsabs.harvard.edu/abs/2025NSRev..12af496L},
}

@Article{HESSCollaboration2024,
  author        = {{H.~E.~S.~S. Collaboration} and Aharonian, F. and Ait Benkhali, F. and Aschersleben, J. and Ashkar, H. and Backes, M. and Barbosa Martins, V. and Batzofin, R. and Becherini, Y. and Berge, D. and Bernl{\"o}hr, K. and Bi, B. and B{\"o}ttcher, M. and Boisson, C. and Bolmont, J. and de Lavergne, M. de Bony and Borowska, J. and Bouyahiaoui, M. and Breuhaus, M. and Brose, R. and Brown, A.~M. and Brun, F. and Bruno, B. and Bulik, T. and Burger-Scheidlin, C. and Caroff, S. and Casanova, S. and Cecil, R. and Celic, J. and Cerruti, M. and Chand, T. and Chandra, S. and Chen, A. and Chibueze, J. and Chibueze, O. and Cotter, G. and Dai, S. and Mbarubucyeye, J. Damascene and Djannati-Ata{\"\i}, A. and Dmytriiev, A. and Doroshenko, V. and Egberts, K. and Einecke, S. and Ernenwein, J.-P. and Filipovic, M. and Fontaine, G. and F{\"u}{\ss}ling, M. and Funk, S. and Gabici, S. and Ghafourizadeh, S. and Giavitto, G. and Glawion, D. and Glicenstein, J.-F. and Grolleron, G. and Haerer, L. and Hinton, J.~A. and Hofmann, W. and Holch, T.~L. and Holler, M. and Horns, D. and Jamrozy, M. and Jankowsky, F. and Jardin-Blicq, A. and Joshi, V. and Jung-Richardt, I. and Kasai, E. and Katarzy{\'n}ski, K. and Khatoon, R. and Kh{\'e}lifi, B. and Klepser, S. and Klu{\'z}niak, W. and Komin, Nu. and Kosack, K. and Kostunin, D. and Kundu, A. and Lang, R.~G. and Le Stum, S. and Leitl, F. and Lemi{\`e}re, A. and Lenain, J.-P. and Leuschner, F. and Lohse, T. and Luashvili, A. and Lypova, I. and Mackey, J. and Malyshev, D. and Malyshev, D. and Marandon, V. and Marchegiani, P. and Marcowith, A. and Mart{\'\i}-Devesa, G. and Marx, R. and Mehta, A. and Mitchell, A. and Moderski, R. and Mohrmann, L. and Montanari, A. and Moulin, E. and Murach, T. and Nakashima, K. and de Naurois, M. and Niemiec, J. and Noel, A. Priyana and Ohm, S. and Olivera-Nieto, L. and de Ona Wilhelmi, E. and Ostrowski, M. and Panny, S. and Panter, M. and Parsons, R.~D. and Peron, G. and Prokhorov, D.~A. and P{\"u}hlhofer, G. and Punch, M. and Quirrenbach, A. and Reichherzer, P. and Reimer, A. and Reimer, O. and Ren, H. and Renaud, M. and Reville, B. and Rieger, F. and Rowell, G. and Rudak, B. and Ricarte, H. Rueda and Ruiz-Velasco, E. and Sahakian, V. and Salzmann, H. and Santangelo, A. and Sasaki, M. and Sch{\"a}fer, J. and Sch{\"u}ssler, F. and Schwanke, U. and Shapopi, J.~N.~S. and Sol, H. and Specovius, A. and Spencer, S. and Stawarz, L. and Steenkamp, R. and Steinmassl, S. and Steppa, C. and Streil, K. and Sushch, I. and Suzuki, H. and Takahashi, T. and Tanaka, T. and Taylor, A.~M. and Terrier, R. and Tsirou, M. and Tsuji, N. and Unbehaun, T. and van Eldik, C. and Vecchi, M. and Veh, J. and Venter, C. and Vink, J. and Wach, T. and Wagner, S.~J. and Werner, F. and White, R. and Wierzcholska, A. and Wong, Yu Wun and Zacharias, M. and Zargaryan, D. and Zdziarski, A.~A. and Zech, A. and Zouari, S. and {\.Z}ywucka, N.},
  journal       = {Science},
  title         = {Acceleration and transport of relativistic electrons in the jets of the microquasar SS 433},
  year          = {2024},
  month         = jan,
  number        = {6681},
  pages         = {402-406},
  volume        = {383},
  archiveprefix = {arXiv},
  doi           = {10.1126/science.adi2048},
  eprint        = {2401.16019},
  primaryclass  = {astro-ph.HE},
  url           = {https://ui.adsabs.harvard.edu/abs/2024Sci...383..402H},
}

@Article{Acharyya2026,
  author        = {{H.~E.~S.~S. Collaboration} and Acharyya, A. and Aharonian, F. and Ashkar, H. and Backes, M. and Batzofin, R. and Berge, D. and Bernl{\"o}hr, K. and B{\"o}ttcher, M. and Boisson, C. and Bolmont, J. and Brun, F. and Bruno, B. and Burger-Scheidlin, C. and Bylund, T. and Casanova, S. and Celic, J. and Cerruti, M. and Chen, A. and Chernyakova, M. and Chibueze, J.~O. and Chibueze, O. and Cornejo, B. and Cotter, G. and de Assis Scarpin, J. and de Bony de Lavergne, M. and de Naurois, M. and de O{\~n}a Wilhelmi, E. and Delgado Giler, A.~G. and Devin, J. and Djannati-Ata{\"\i}, A. and Dmytriiev, A. and Egberts, K. and Egg, K. and Ernenwein, J.-P. and Esca{\~n}uela Nieves, C. and Fauverge, P. and Feijen, K. and Filipovic, M.~D. and Fontaine, G. and Funk, S. and Gabici, S. and Gallant, Y.~A. and Glicenstein, J.~F. and Glombitza, J. and Goswami, P. and Grondin, M.-H. and Heckmann, L. and He{\ss}, B. and Hinton, J.~A. and Hofmann, W. and Holch, T.~L. and Holler, M. and Jamrozy, M. and Jankowsky, F. and Jardin-Blicq, A. and Jaroschewski, I. and Jimeno, D. and Jung-Richardt, I. and Katarzy{\'n}ski, K. and Kerszberg, D. and Kh{\'e}lifi, B. and Komin, N. and Kosack, K. and Kostunin, D. and Lang, R.~G. and Lazarevi{\'c}, S. and Lemi{\`e}re, A. and Lemoine-Goumard, M. and Lenain, J.-P. and Liniewicz, P. and Luashvili, A. and Mackey, J. and Malyshev, D. and Marandon, V. and Mayer, M.~G.~F. and Mehta, A. and Mitchell, A.~M.~W. and Moderski, R. and Mohrmann, L. and Montanari, A. and Moulin, E. and Niemiec, J. and Olivera-Nieto, L. and Moghadam, M.~O. and Panny, S. and Parsons, R.~D. and Pensec, U. and Pichard, P. and Preis, T. and P{\"u}hlhofer, G. and Punch, M. and Quirrenbach, A. and Reimer, A. and Reimer, O. and Reis, I. and Remy, Q. and Ren, H.~X. and Reville, B. and Rieger, F. and Roellinghoff, G. and Rowell, G. and Rudak, B. and Sabri, K. and Safi-Harb, S. and Sahakian, V. and Santangelo, A. and Sasaki, M. and Sch{\"u}ssler, F. and Shapopi, J.~N.~S. and Si Said, W. and Sol, H. and Stawarz, {\L}. and Steinmassl, S. and Tanaka, T. and Taylor, A.~M. and Taylor, G.~L. and Terrier, R. and Tian, Y. and Timmermans, A. and Tsirou, M. and Tsuji, N. and Unbehaun, T. and van Eldik, C. and Vecchi, M. and Venter, C. and Vink, J. and Voitsekhovskyi, V. and Wagner, S.~J. and Wierzcholska, A. and Zacharias, M. and Zdziarski, A.~A. and Zech, A. and Zhong, W. and Takekawa, S.},
  journal       = {\aap},
  title         = {Constraining the nature of the most extreme Galactic particle accelerator: H.E.S.S. observations of the microquasar V4641 Sgr},
  year          = {2026},
  month         = jan,
  pages         = {A8},
  volume        = {706},
  archiveprefix = {arXiv},
  doi           = {10.1051/0004-6361/202557532},
  eid           = {A8},
  eprint        = {2511.10537},
  primaryclass  = {astro-ph.HE},
  url           = {https://ui.adsabs.harvard.edu/abs/2026A&A...706A...8A},
}

@Article{Abeysekara2018,
  author        = {{HAWC Collaboration} and Abeysekara, A.~U. and Albert, A. and Alfaro, R. and Alvarez, C. and {\'A}lvarez, J.~D. and Arceo, R. and Arteaga-Vel{\'a}zquez, J.~C. and Avila Rojas, D. and Ayala Solares, H.~A. and Belmont-Moreno, E. and BenZvi, S.~Y. and Brisbois, C. and Caballero-Mora, K.~S. and Capistr{\'a}n, T. and Carrami{\~n}ana, A. and Casanova, S. and Castillo, M. and Cotti, U. and Cotzomi, J. and Couti{\~n}o de Le{\'o}n, S. and De Le{\'o}n, C. and De la Fuente, E. and D{\'\i}az-V{\'e}lez, J.~C. and Dichiara, S. and Dingus, B.~L. and DuVernois, M.~A. and Ellsworth, R.~W. and Engel, K. and Espinoza, C. and Fang, K. and Fleischhack, H. and Fraija, N. and Galv{\'a}n-G{\'a}mez, A. and Garc{\'\i}a-Gonz{\'a}lez, J.~A. and Garfias, F. and Gonz{\'a}lez-Mu{\~n}oz, A. and Gonz{\'a}lez, M.~M. and Goodman, J.~A. and Hampel-Arias, Z. and Harding, J.~P. and Hernandez, S. and Hinton, J. and Hona, B. and Hueyotl-Zahuantitla, F. and Hui, C.~M. and H{\"u}ntemeyer, P. and Iriarte, A. and Jardin-Blicq, A. and Joshi, V. and Kaufmann, S. and Kar, P. and Kunde, G.~J. and Lauer, R.~J. and Lee, W.~H. and Le{\'o}n Vargas, H. and Li, H. and Linnemann, J.~T. and Longinotti, A.~L. and Luis-Raya, G. and L{\'o}pez-Coto, R. and Malone, K. and Marinelli, S.~S. and Martinez, O. and Martinez-Castellanos, I. and Mart{\'\i}nez-Castro, J. and Matthews, J.~A. and Miranda-Romagnoli, P. and Moreno, E. and Mostaf{\'a}, M. and Nayerhoda, A. and Nellen, L. and Newbold, M. and Nisa, M.~U. and Noriega-Papaqui, R. and Pretz, J. and P{\'e}rez-P{\'e}rez, E.~G. and Ren, Z. and Rho, C.~D. and Rivi{\`e}re, C. and Rosa-Gonz{\'a}lez, D. and Rosenberg, M. and Ruiz-Velasco, E. and Salesa Greus, F. and Sandoval, A. and Schneider, M. and Schoorlemmer, H. and Seglar Arroyo, M. and Sinnis, G. and Smith, A.~J. and Springer, R.~W. and Surajbali, P. and Taboada, I. and Tibolla, O. and Tollefson, K. and Torres, I. and Vianello, G. and Villase{\~n}or, L. and Weisgarber, T. and Werner, F. and Westerhoff, S. and Wood, J. and Yapici, T. and Yodh, G. and Zepeda, A. and Zhang, H. and Zhou, H.},
  journal       = {\nat},
  title         = {Very-high-energy particle acceleration powered by the jets of the microquasar SS 433},
  year          = {2018},
  month         = oct,
  number        = {7725},
  pages         = {82-85},
  volume        = {562},
  archiveprefix = {arXiv},
  doi           = {10.1038/s41586-018-0565-5},
  eprint        = {1810.01892},
  primaryclass  = {astro-ph.HE},
  url           = {https://ui.adsabs.harvard.edu/abs/2018Natur.562...82A},
}

@Article{Alfaro2024,
  author        = {{HAWC Collaboration} and Alfaro, R. and Alvarez, C. and Arteaga-Vel{\'a}zquez, J.~C. and Avila Rojas, D. and Ayala Solares, H.~A. and Babu, R. and Belmont-Moreno, E. and Caballero-Mora, K.~S. and Capistr{\'a}n, T. and Carrami{\~n}ana, A. and Casanova, S. and Cotti, U. and Cotzomi, J. and Couti{\~n}o de Le{\'o}n, S. and De la Fuente, E. and Depaoli, D. and Di Lalla, N. and Diaz Hernandez, R. and Dingus, B.~L. and DuVernois, M.~A. and Durocher, M. and D{\'\i}az-V{\'e}lez, J.~C. and Engel, K. and Espinoza, C. and Fan, K.~L. and Fang, K. and Fraija, N. and Fraija, S. and Garc{\'\i}a-Gonz{\'a}lez, J.~A. and Garfias, F. and Gonzalez Mu{\~n}oz, A. and Gonz{\'a}lez, M.~M. and Goodman, J.~A. and Groetsch, S. and Harding, J.~P. and Herzog, I. and Hinton, J. and Huang, D. and Hueyotl-Zahuantitla, F. and H{\"u}ntemeyer, P. and Iriarte, A. and Joshi, V. and Kaufmann, S. and Kieda, D. and de Le{\'o}n, C. and Lee, J. and Le{\'o}n Vargas, H. and Linnemann, J.~T. and Longinotti, A.~L. and Luis-Raya, G. and Malone, K. and Martinez, O. and Mart{\'\i}nez-Castro, J. and Matthews, J.~A. and Miranda-Romagnoli, P. and Morales-Soto, J.~A. and Moreno, E. and Mostaf{\'a}, M. and Nayerhoda, A. and Nellen, L. and Newbold, M. and Nisa, M.~U. and Noriega-Papaqui, R. and Olivera-Nieto, L. and Omodei, N. and Osorio, M. and P{\'e}rez Araujo, Y. and P{\'e}rez-P{\'e}rez, E.~G. and Rho, C.~D. and Rosa-Gonz{\'a}lez, D. and Ruiz-Velasco, E. and Salazar, H. and Salazar-Gallegos, D. and Sandoval, A. and Schneider, M. and Serna-Franco, J. and Smith, A.~J. and Son, Y. and Springer, R.~W. and Tibolla, O. and Tollefson, K. and Torres, I. and Torres-Escobedo, R. and Turner, R. and Ure{\~n}a-Mena, F. and Varela, E. and Villase{\~n}or, L. and Wang, X. and Watson, I.~J. and Willox, E. and Yun-C{\'a}rcamo, S. and Zhou, H.},
  journal       = {\nat},
  title         = {Ultra-high-energy gamma-ray bubble around microquasar V4641 Sgr},
  year          = {2024},
  month         = oct,
  number        = {8034},
  pages         = {557-560},
  volume        = {634},
  archiveprefix = {arXiv},
  doi           = {10.1038/s41586-024-07995-9},
  eprint        = {2410.16117},
  primaryclass  = {astro-ph.HE},
  url           = {https://ui.adsabs.harvard.edu/abs/2024Natur.634..557A},
}

@Article{Popescu2017,
  author        = {Popescu, C.~C. and Yang, R. and Tuffs, R.~J. and Natale, G. and Rushton, M. and Aharonian, F.},
  journal       = {\mnras},
  title         = {A radiation transfer model for the Milky Way: I. Radiation fields and application to high-energy astrophysics★},
  year          = {2017},
  month         = sep,
  number        = {3},
  pages         = {2539-2558},
  volume        = {470},
  archiveprefix = {arXiv},
  doi           = {10.1093/mnras/stx1282},
  eprint        = {1705.06652},
  primaryclass  = {astro-ph.GA},
  url           = {https://ui.adsabs.harvard.edu/abs/2017MNRAS.470.2539P},
}

@article{Goranskij2003,
    title = {Photometry and spectroscopy of the system {V4641} sagittarii in quiescence},
    volume = {47},
    url = {https://ui.adsabs.harvard.edu/abs/2003ARep...47..740G},
    doi = {10.1134/1.1611215},
    number = {9},
    journal = {Astronomy Reports},
    author = {Goranskij, V. P. and Barsukova, E. A. and Burenkov, A. N.},
    month = sep,
    year = {2003},
    pages = {740--749},
}

@article{Orosz2001,
    title = {A black hole in the superluminal source {SAX} {J1819}.3-2525 ({V4641} sgr)},
    volume = {555},
    url = {https://ui.adsabs.harvard.edu/abs/2001ApJ...555..489O},
    doi = {10.1086/321442},
    number = {1},
    journal = {ApJ},
    author = {Orosz, Jerome A. and Kuulkers, Erik and van der Klis, Michiel and McClintock, Jeffrey E. and Garcia, Michael R. and Callanan, Paul J. and Bailyn, Charles D. and Jain, Raj K. and Remillard, Ronald A.},
    month = jul,
    year = {2001},
    pages = {489--503},
}

@article{MacDonald2014,
    title = {The black hole binary {V4641} sagitarii: {Activity} in quiescence and improved mass determinations},
    volume = {784},
    url = {https://ui.adsabs.harvard.edu/abs/2014ApJ...784....2M},
    doi = {10.1088/0004-637X/784/1/2},
    number = {1},
    journal = {ApJ},
    author = {MacDonald, Rachel K. D. and Bailyn, Charles D. and Buxton, Michelle and Cantrell, Andrew G. and Chatterjee, Ritaban and Kennedy-Shaffer, Ross and Orosz, Jerome A. and Markwardt, Craig B. and Swank, Jean H.},
    month = mar,
    year = {2014},
    pages = {2},
}

@article{Markwardt1999,
    title = {Xte j1819-254, xte j1743-363, xte j1710-281, xte j1723-376},
    volume = {7120},
    journal = {IAU Circ},
    author = {Markwardt, C. B. and Swank, J. H. and Marshall, F. E.},
    month = mar,
    year = {1999},

    pages = {1},
}

@article{Zand2000,
    title = {{BeppoSAX} observations of the nearby low-mass {X}-ray binary and fast transient {SAX} {J1819}.3-2525},
    volume = {357},
    url = {https://ui.adsabs.harvard.edu/abs/2000A&A...357..520I},
    doi = {10.48550/arXiv.astro-ph/0001110},
    journal = {A\&A},
    author = {in't Zand, J. J. M. and Kuulkers, E. and Bazzano, A. and Cornelisse, R. and Cocchi, M. and Heise, J. and Muller, J. M. and Natalucci, L. and Smith, M. J. S. and Ubertini, P.},
    month = may,
    year = {2000},
    pages = {520--526},
}

@article{Wijnands2000,
    title = {The rapid x-ray variability of {V4641} sagittarii ({SAX} {J1819}.3-2525 = {XTE} {J1819}-254)},
    volume = {528},
    url = {https://ui.adsabs.harvard.edu/abs/2000ApJ...528L..93W},
    doi = {10.1086/312439},
    number = {2},
    journal = {ApJL},
    author = {Wijnands, Rudy and van der Klis, Michiel},
    month = jan,
    year = {2000},
    pages = {L93--L96},
}

@article{Shaw2022,
    title = {High resolution {X}-ray spectroscopy of {V4641} {Sgr} during its 2020 outburst},
    volume = {516},
    url = {https://ui.adsabs.harvard.edu/abs/2022MNRAS.516..124S},
    doi = {10.1093/mnras/stac2213},
    number = {1},
    journal = {MNRAS},
    author = {Shaw, A. W. and Miller, J. M. and Grinberg, V. and Buisson, D. J. K. and Heinke, C. O. and Plotkin, R. M. and Tomsick, J. A. and Bahramian, A. and Gandhi, P. and Sivakoff, G. R.},
    month = oct,
    year = {2022},
    pages = {124--137},
}

@article{Hjellming2000,
    title = {Light curves and radio structure of the 1999 september transient event in {V4641} sagittarii (={XTE} {J1819}-254={SAX} {J1819}.3-2525)},
    volume = {544},
    url = {https://ui.adsabs.harvard.edu/abs/2000ApJ...544..977H},
    doi = {10.1086/317255},
    number = {2},
    journal = {ApJ},
    author = {Hjellming, R. M. and Rupen, M. P. and Hunstead, R. W. and Campbell-Wilson, D. and Mioduszewski, A. J. and Gaensler, B. M. and Smith, D. A. and Sault, R. J. and Fender, R. P. and Spencer, R. E. and de la Force, C. J. and Richards, A. M. S. and Garrington, S. T. and Trushkin, S. A. and Ghigo, F. D. and Waltman, E. B. and McCollough, M.},
    month = dec,
    year = {2000},
    pages = {977--992},
}

@article{Crook-Mansour2026,
       author = {{Crook-Mansour}, Justine and {Fender}, Rob and {Hughes}, Andrew and {Motta}, Sara and {Woudt}, Patrick A. and {Bahramian}, Arash and {Del Santo}, Melania and {Zhang}, Zuobin and {Russell}, Thomas D. and {van den Eijnden}, Jakob and {Bright}, Joe and {Williams-Baldwin}, David and {Carotenuto}, Francesco and {Corbel}, St{\'e}phane and {Cowie}, Fraser J. and {Andersson}, Alex and {Grollimund}, Noa and {Matthews}, James and {Gasealahwe}, Kelebogile and {Monageng}, Itumeleng and {Rhodes}, Lauren and {Saikia}, Payaswini and {Savard}, Katie and {Tremou}, Evangelia and {Zhang}, Xian},
        title = "{The homogeneous MeerKAT and Swift/XRT X-ray binary radio:X-ray plane}",
      journal = {\mnras},
         year = 2026,
        month = sep,
       volume = {551},
       number = {2},
          eid = {stag1022},
        pages = {stag1022},
          doi = {10.1093/mnras/stag1022},
archivePrefix = {arXiv},
       eprint = {2605.14003},
 primaryClass = {astro-ph.HE},
       adsurl = {https://ui.adsabs.harvard.edu/abs/2026MNRAS.551g1022C}
}

@article{Wan2026,
  title = {A {{Leptonic Interpretation}} of the {{Ultra-high-energy Gamma-Ray Emission}} from {{V4641 Sgr}}},
  author = {Wan, Su-Yu and Wang, Jie-shuang and Liu, Ruo-Yu},
  year = 2026,
  month = jan,
  journal = {ApJ},
  volume = {997},
  number = {1},
  pages = {128},
  issn = {0004-637X, 1538-4357},
  doi = {10.3847/1538-4357/ae246c},
  urldate = {2026-05-19}
}

@article{Kawamuro2018,
    title = {{MAXI}/{GSC} detection of a probable new {X}-ray transient {MAXI} {J1820}+070},
    volume = {11399},
    url = {https://ui.adsabs.harvard.edu/abs/2018ATel11399....1K},
    urldate = {2026-05-19},
    journal = {The Astronomer's Telegram},
    author = {Kawamuro, T. and Negoro, H. and Yoneyama, T. and Ueno, S. and Tomida, H. and Ishikawa, M. and Sugawara, Y. and Isobe, N. and Shimomukai, R. and Mihara, T. and Sugizaki, M. and Nakahira, S. and Iwakiri, W. and Yatabe, F. and Takao, Y. and Matsuoka, M. and Kawai, N. and Sugita, S. and Yoshii, T. and Tachibana, Y. and Harita, S. and Morita, K. and Yoshida, A. and Sakamoto, T. and Serino, M. and Kawakubo, Y. and Kitaoka, Y. and Hashimoto, T. and Tsunemi, H. and Nakajima, M. and Kawase, T. and Sakamaki, A. and Maruyama, W. and Ueda, Y. and Hori, T. and Tanimoto, A. and Oda, S. and Morita, T. and Yamada, S. and Tsuboi, Y. and Nakamura, Y. and Sasaki, R. and Kawai, H. and Sato, T. and Yamauchi, M. and Hanyu, C. and Hidaka, K. and Yamaoka, K. and Shidatsu, M.},
    month = mar,
    year = {2018},
    pages = {1},
}

@article{Kaci2025,
       author = {{Kaci}, Samy and {Giacinti}, Gwenael and {Aharonian}, Felix and {Wang}, Jie-Shuang},
        title = "{Microquasars as the major contributors to Galactic cosmic rays around the ``knee''}",
      journal = {arXiv e-prints},
         year = 2025,
        month = oct,
          eid = {arXiv:2510.01369},
        pages = {arXiv:2510.01369},
          doi = {10.48550/arXiv.2510.01369},
archivePrefix = {arXiv},
       eprint = {2510.01369},
 primaryClass = {astro-ph.HE},
       adsurl = {https://ui.adsabs.harvard.edu/abs/2025arXiv251001369K}
}

\appendix
\section{Optical counterpart and extinction}
\label{app:opt}

\begin{figure}
    \centering
    
        \includegraphics[width=0.5\textwidth]{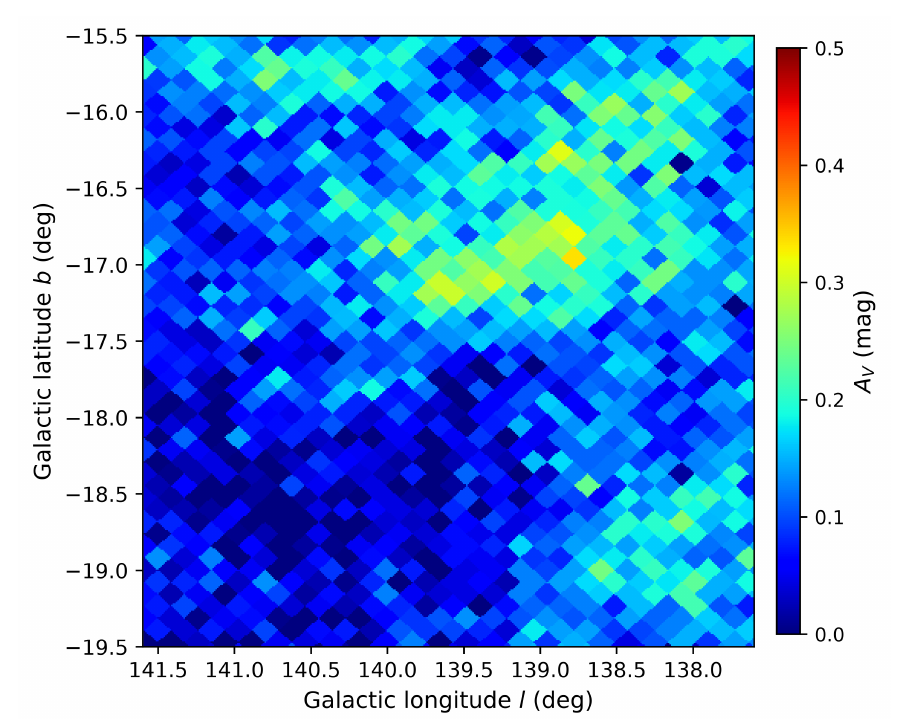} 

    \caption{\label{fig: extinction} The extinction $A_V$ in the direction of the expected position of an \xray binary at the centre of the peanut (for a distance of $1.3$ kpc)}
\end{figure}

Given its location off the Galactic Plane, the optical extinction in the direction of the peanut is very low. To estimate the extinction, we use the \textsc{dustmaps} package in \textsc{python} \citep{2018JOSS....3..695M}. We specifically use the map from \citet{2019ApJ...887...93G} to study the field around $(l,b) = (139.6,-17.5)$. Along that direction, we measure $A_V = 0.15$ at a distance of $1.3$ kpc (or beyond). In Figure \ref{fig: extinction}, we show the inferred $A_V$ across the field of view around this position. Regardless of its exact position, the expected extinction to the \xray binary is negligible.

Consequently, if the \xray binary hosted a high-mass star, extinction would not prevent its identification. The third edition of the Alma catalogue of OB stars does not include any massive stars or candidates in the potential location region of the \xray binary \citep{2025MNRAS.543...63P}: the closest one is located at $(l,b) = (140.78, -18.2)$. Similarly, \textit{Gaia} DR3, queried, filtered, and searched following the same approach as \citep{2024A&A...681A..21S}, does not reveal any massive star candidates missed by the Alma catalogue. Therefore, we conclude that the proposed \xray counterpart must then be a low-mass \xray binary. This identification fits with the smaller scale height of low-mass \xray binaries than high-mass \xray binaries \citep{2023A&A...677A.134N,2023A&A...675A.199A}, and the low scale height of massive stars in general \citep{2001A&A...378..907R}.

\begin{table}[b]
 \caption[]{\label{tab:params} Values of the parameters used in Figure~\ref{fig:spec}.}
\resizebox{\columnwidth}{!}{\begin{tabular}{lcc}
 \hline \hline
 \noalign{\vskip 1mm} 

  Parameter & Meaning &  Value   \\
\noalign{\vskip 1mm} 

\hline
\noalign{\smallskip}
\multicolumn{3}{c}{ leptonic model, $t_{\rm inj}=5\times10^3$~yr} \\

\noalign{\smallskip}
\hline
\noalign{\vskip 1mm} 
$B$ ($\mu G$)   &  magnetic field &2.87  \\
$\alpha_e$ (\% of L$_{\mathrm{Edd}}$)   & normalization parameter & 9$\cdot 10^{-4}$ \\
$\Gamma_e$   &  electron spectral index & 1.93 \\
$E_{\rm cut}^{e}$ (PeV) & electron exponential cutoff energy  &  0.99 \\
\noalign{\vskip 1mm} 
\hline

\end{tabular}}
\tablefoot{All parameter values except $t_{\rm inj}$ and $\alpha_e$ are taken directly from~\citetalias{Acharyya2026}. The normalization parameter $\alpha_e$ parametrizes the electron luminosity L$_e$ as a percentage of L$_{\mathrm{Edd}}$.}
\end{table}
\end{document}